# Two-dimensional Janus van der Waals heterojunctions: a review of recent research progresses

Lin Ju[1,2], Mei Bie[3], Xiwei Zhang[2], Xiangming Chen[2], and Liangzhi Kou[1]*

[1]School of Mechanical, Medical and Process Engineering, Queensland University of Technology, Gardens Point Campus, QLD 4001, Brisbane, Australia

[2]School of Physics and Electric Engineering, Anyang Normal University, Anyang, 455000, China

[3]Shandong Institute for Food and Drug Control, Jinan, 250101, China

Corresponding Email: liangzhi.kou@qut.edu.au

**Abstract**

Two-dimensional Janus van der Waals (vdW) heterojunctions, referring to the junction containing at least one Janus material, are found to exhibit tuneable electronic structures, wide light adsorption spectra, controllable contact resistance, and sufficient redox potential due to the intrinsic polarization and unique interlayer coupling. These novel structures and properties are promising for the potential applications in electronics and energy conversion devices. To provide a comprehensive picture about the research progress and guide the following investigations, here we summarize their fundamental properties of different types of two-dimensional Janus vdW heterostructures including electronic structure, interface contact and optical properties, and discuss the potential applications in electronics and energy conversion devices. The further challenges and possible research directions of the novel heterojunctions are discussed at the end of this review.

# 1. Introduction

In recent years, two-dimensional (2D) layered materials, with graphene [1, 2], graphitic carbon nitride (g-$C_3N_4$) [3, 4], molybdenum disulphide ($MoS_2$) [5, 6], phosphorene [7, 8], and MXenes [9] as the representative examples, have attracted extensive research interests due to the excellent electronic/mechanical properties and potential application in nanodevices. Different from single materials, vdW heterostructure that combines two or more different layered materials creates more opportunities for novel properties and potential applications [10, 11]. Due to the relative weak interlayer coupling, most of the intrinsic properties from the components remain, which give the opportunities to combine the intrinsic advantages in the heterostructures. Besides, the vdW heterostructure can also overcome the shortcomings of single materials, such as low quantum efficiency, high charge recombination, and serious chemical back-reactions [12, 13]. As a typical example, the redox potential *vs.* light absorption in the semiconductor is an irreconcilable contradiction for photocatalytic water-splitting. A small band gap is beneficial for the high light absorption, in contrast a large band gap is essential for a high redox potential for water splitting [14]. The paradox is hard to be solved in single material but can be addressed in vdW heterostructures by combining the layers with different band gaps together. It is also interesting to notice that, the new phenomena can be induced as a result of the interlayer polarization in 2D vdW heterojunction, such as new optical absorption peak [15], charge migration [16], built-in field [17], and rearranged band alignments [18]. Various nanodevices based on the heterostructures have been demonstrated or proposed, such as field-effect transistors [19, 20], solar cells [21, 22], light emitting diode (LED) [23], and photodetectors [24, 25].

Among the 2D vdW heterostructures, a special family has attracted more and more research attentions, namely the multilayer junction composed with at least one layer of Janus material (in the following, it is short for 2D Janus vdW heterostructure). Different from traditional vdW heterostructures, the intrinsic intralayer polarization from Janus material will couple with the interlayer built-in polarization field, which will provide an additional degree of freedom to modulate the physical/chemical property of the heterostructure, leading to the novel features and potentials for applications. Based on the recent research progresses, in the present review, after briefly introducing the theoretical stability of Janus vdW heterostructures, we then comprehensively summarize their fundamental electronic/optical/chemical properties (see **Fig. 1**). The associated potential applications in specific areas like electronic devices, chemical catalysis and energy conversion are also discussed. The review is expected to provide a comprehensive overview of the special heterostructure family and inspire the experimental demonstrations in the near future. At the end of the review, the further challenges and possible research directions are addressed.

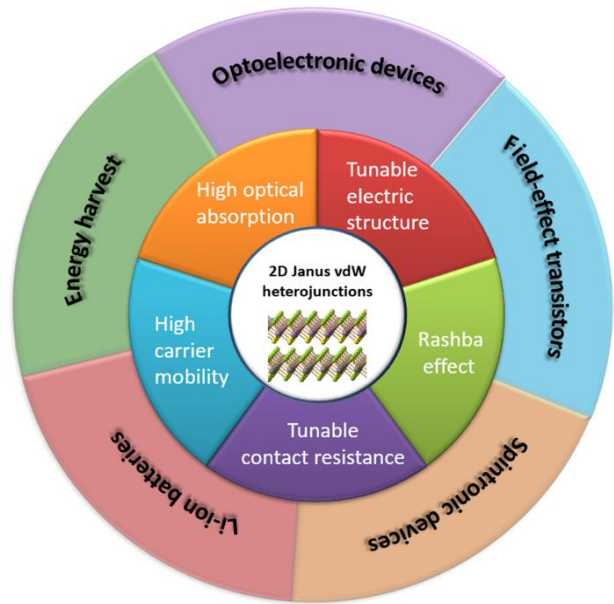

**Fig. 1** Fundamental properties and potential applications of 2D Janus vdW heterojunctions.

## 2. Fabrication of Janus materials and stability of potential Janus vdW heterostructures

As essential components of Janus vdW heterostructures, until now there are only two kinds of synthesized Janus layer materials, namely MoSSe and WSSe monolayers. Janus WSSe monolayer was fabricated by implanting Se species into $WS_2$ monolayer with pulsed laser ablation plasmas [26], while MoSSe monolayer was synthesized based on the modified chemical vapor deposition (CVD) methods of selenization of $MoS_2$ monolayer [27] or sulfurization of $MoSe_2$ monolayer [28], see **Figs. 2a** and **2b**. The selective sulfurization process has a serve requirement for the temperature and pressure (see **Fig. 2b**). Under a maintained atmosphere pressure, when the temperature is below 750 °C and over 850 °C, the Raman peaks of the samples are similar to the ones of pristine $MoSe_2$ and $MoS_2$, respectively. Only when the temperature is between 750 °C and 850 °C, the selenium substitution reaction occurs at the top layer, but not at the bottom layer, leading to the successful synthesis of Janus MoSSe monolayer. It also has been found that the atmospheric pressure could enlarge the stable temperature window for the selenium substitution reaction at the top layer, but the long sulfurization time could not trigger the selenium substitution reaction at the bottom layer.

For the fabrication of vertical heterostructures, there are mainly two strategies, either by manually stacking different exfoliated nanosheets [29-31] or directly growing different nanosheets on the selected substrate [22, 32-34]. For the first method, nanosheets such as boron nanosheet [35], antimonene [36], bismuthene [37], few-layered InSe [38], and $CH_3NH_3PbI_3$ perovskite nanosheets [39] need to be physically transferred to the substrates. However, the major drawback is that, the procedure of chemical intercalation or mechanical sonication process is neither scalable nor controllable [40]. For the second strategy of CVD technique, the fabricated heterojunction interface is clean, and the stacking is controllable [32].

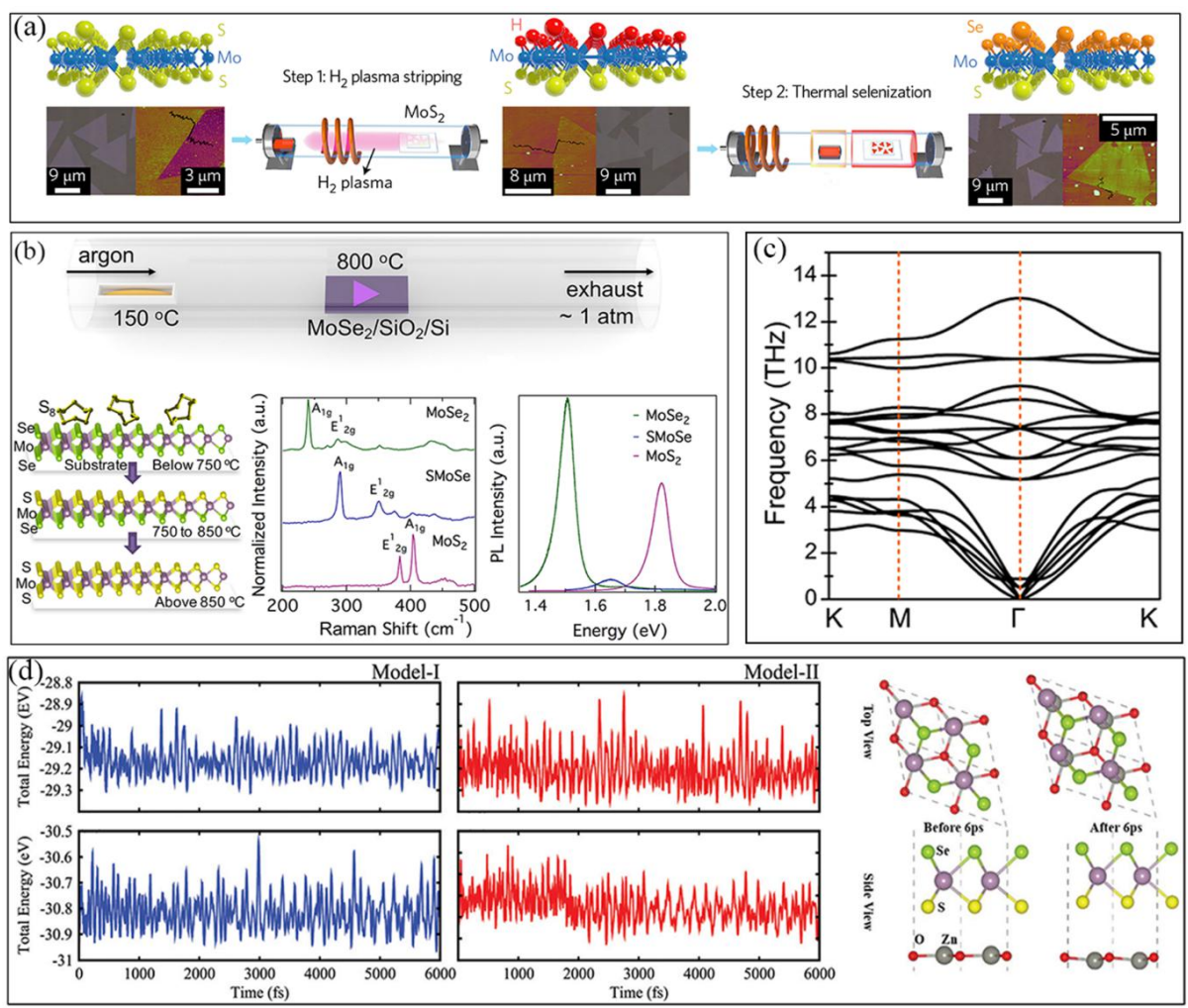

**Fig. 2** (a), (b) Schematic illustration of the reaction setup for the fabrication of Janus MoSSe monolayer with the two different CVD methods. (c) Phonon dispersion of MoSSe-WSe$_2$ vdW heterojunction. (d) (left)Thermal stabilities of MoSSe–ZnO (up) and WSSe–ZnO (down) vdW heterostructures with different stacking patterns, and (right) ZnO-MoSSe atomic structure before and after heating for 6 ps. (a) Reproduced from Ref.[27]. (b) Reproduced from Ref.[28]. (c) Reproduced from Ref.[41]. (d) Reproduced from Ref.[42].

Although there is no experimental report of the 2D Janus vdW heterostructures until now, the structural stability as a function of stacking patterns have been comprehensively investigated in numerous theoretical works, from the energetic, dynamical, thermal and mechanical perspectives. They were evaluated by the binding energy, phonon spectrum, molecular dynamics, and elastic constants respectively. For example, for VSe$_2$-MoSSe, MoSSe-WSe$_2$, MoSSe-WSSe, GeC-MoSSe, GeC-WSSe, and BlueP (blue phosphorus)-MoSSe vdW heterostructures, [42-49] the calculated binding energies are all negative, which confirm the energetic stability of these 2D Janus vdW heterostructures. Besides, the dynamical stabilities were verified by the phonon calculations. Specifically, the imaginary frequency is absent throughout the Brillouin zone with suitable stacking pattern [41, 43, 44, 47, 50, 51] (see the phonon spectrum of MoSSe-WSe$_2$ vdW heterostructure as an example in **Fig. 2c**). For the supercells of MoSSe-WSe$_2$, ZnO-MoSSe, ZnO-WSSe, MoSSe-GaN, MoSSe-AlN, and BlueP-MoSSe vdW heterostructures, their thermal stabilities were checked with the *ab initio* molecular dynamics (AIMD) simulations at 300 K for several picoseconds. The slight total energy fluctuation and insignificant geometric reconstructions demonstrate their good thermal stability at room temperature [41, 42, 46, 48], (see **Fig. 2d**, for ZnO-MoSSe vdW heterostructure). The mechanical stability of MoSSe-WSSe and GaS$_{0.5}$Se$_{0.5}$-Arsenene vdW heterostructures has been demonstrated based on Born's stability criteria ( $C_{ii} > 0$, and $C_{11}C_{22} - C_{12}^2 > 0$ ) [43, 49]. All these theoretical investigations about stabilities combined with the fact that some Janus layered materials have been experimentally synthesized, indicate that the 2D Janus vdW heterostructure is stable and feasible to be fabricated in the near future.

## 3. Fundamental properties

### 3.1. Electronic structure

Due to the presence of intrinsic intralayer polarization from Janus materials and its coupling with interlayer polarization field, the electronic properties of 2D Janus vdW heterojunction will be significantly affected. The recent researches indicate that the band gap, band alignment, band edge positions, and band splitting will strongly depend on the stacking orders, external electric fields and mechanical strains.

## Band gap engineering under strain and electric field

As a result of the intralayer, interlayer polarization and the strain induced by the lattice mismatch, the band gap of 2D Janus vdW heterostructures is susceptible to the external disturbance like strain and electric field. Take MoSSe-WSSe vdW heterostructure as an example, the band gaps can be significantly reduced with the external in-plane strains, as illustrated in **Fig. 3a**. When the tensile strain is larger than 8% or a large electric field is applied, the heterostructure system can be metallized [50]. Under in-plane biaxial strains or vertical compressive strain (by changing the interlayer distance, see **Fig. 3c**), a direct–indirect transition of band gap can be induced. However, the phenomenon is absent in vertical tensile strain, since the coupling between interlayers will be weakened when the distance is increased [43]. For the $In_2STe$-InSe vdW heterostructure, the characteristic of intrinsic direct-bandgap could be retained under vertical tensile strain, although the value becomes smaller. Whereas, the vertical electric field gives rise to a direct-indirect transformation of band structure, see **Fig. 3c** [52]. Very similar band gap behaviour under strain or electric field can be found in WSSe-SiC, MoSSe-SiC, MoSTe–WSTe, MoSeTe–WSeTe, and MoSSe-GaN vdW heterostructures [45, 53-55], indicating that both the external strain and electric field are effective approaches to modulate the electronic properties of 2D Janus vdW heterostructures.

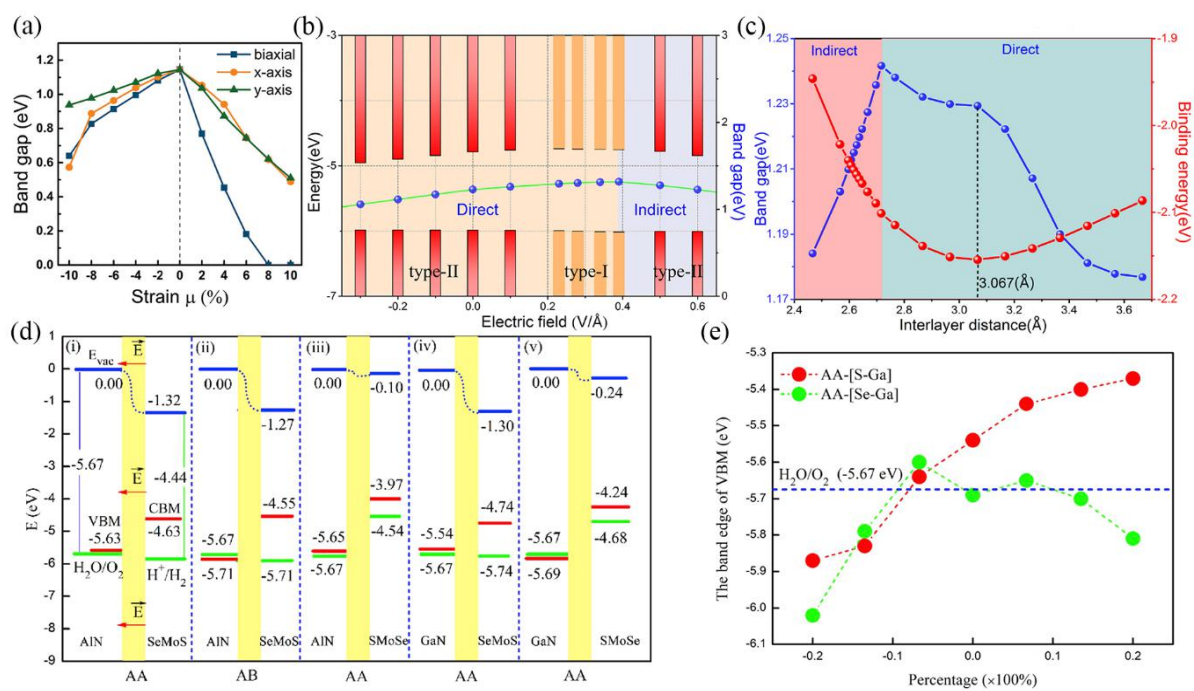

**Fig. 3** (a) The calculated bandgap–strain relationships of MoSSe-WSSe vdW heterostructure under different uniaxial and biaxial strains. (b) The band gap and band edge location of $In_2STe$-InSe vdW heterostructure with different electric fields. (c) The binding energies and band gaps of In$_2$STe-InSe vdW heterostructure versus the interlayer distance. (d) The band edge positions of MoSSe-AlN and MoSSe-GaN vertical heterostructures with different stacking patterns. (e) VBM of MoSSe-GaN vertical heterostructures with different stacking patterns under different strains. (a) Reproduced from Ref.[43]. (b), (c) Reproduced from Ref.[52]. (d), (e) Reproduced from Ref.[56].

**Band alignment and band edge positions**

As an important electronic parameter of heterostructures, the band alignment between the two components will determine the optical adsorption and charge separation, which is highly related with photovoltaic and photocatalytic application. Different from the traditional heterostructures, the band alignment of Janus vdW heterostructures can be tuned by the intrinsic polarization of Janus layer, where the direction can be adjusted by the stacking order. For example, when the stacking order is changed from BuleP-S-Mo-Se to BlueP-Se-Mo-S, the band alignment of BlueP-MoSSe vdW heterostructure turns from type-I into type-II as desirable for optical usage [48]. Similar transition is also observed in MoSSe-WS$_2$ and MoSSe-WSe$_2$ heterostructures with the aid of large additional electric field and strains [55]. As displayed in **Fig. 3b**, the transition of staggered-straddling band alignment can be triggered for the In$_2$STe-InSe vdW heterostructure, under either external electric field or vertical strains [52].

For the type II heterostructure, the conduction band minimum (CBM) and valance band maximum (VBM) locate at different components. Moreover, due to the different electronegativity, the components have different vacuum potentials [17, 18]. Hence, the band edge positions, the reduction potential of H$^+$/H$_2$, and the oxidation potential of O$_2$/H$_2$O, which are normally used to evaluate the redox capacities of carriers for water-splitting, need to be calculated separately in the different components. Unfortunately, this point has been ignored by some of the studies, leading to unreliable predictions [45, 57, 58]. Even so, the band edge positions of the MoSSe-AlN and MoSSe-GaN vdW heterostructures were appropriately evaluated (shown in **Fig. 3d**), it is predicted to be suitable for water-splitting. Meanwhile it is found that the band edge positions could be tuned by external electric field, strains or stacking patterns, see **Fig. 3e** [53, 56]. Furthermore, utilizing the variations of band edge positions and total energy with respect to strain, the carrier mobility could be calculated based on the Deformation Potential theory (the detailed calculated method can be found in our previous article [59]). For the MoSSe/GaN and MoSSe/AlN vdW heterostructures, the electron mobility, along the armchair direction, is 275.86 and 384.51 cm$^2 \cdot$V$^{-1} \cdot$s$^{-1}$, and along the zigzag direction, is 276.27 and 575.08 cm$^2 \cdot$V$^{-1} \cdot$s$^{-1}$, respectively. Meanwhile, the hole mobility, along the armchair direction, is 3476.81 and 280.27 cm$^2 \cdot$V$^{-1} \cdot$s$^{-1}$, and along the zigzag direction, is 3651.83 and 334.11 cm$^2 \cdot$V$^{-1} \cdot$s$^{-1}$, respectively [46]. Compared with the carrier mobility of MoSSe monolayer (210.95 and 52.72 cm$^2 \cdot$V$^{-1} \cdot$S$^{-1}$ for hole and electron mobility) [60], the formation of heterostructures speeds up the carrier transfer [46].

**Rashba effect**

Due to the intrinsic polarization in a Janus material, the symmetry along out-of-plane direction for 2D Janus vdW heterostructure is absent. When the spin orbit coupling is considered, the intralayer and interlayer polarizations of the heterostructure will not only induce band gap variation and band edge position shifts, but also lead to spin band splitting and shifts in the reciprocal space, namely Rashba effects. Similar to the particles/anti-particles splitting in Dirac Hamiltonian, the Rashba effect is momentum dependent spin bands splitting, which is a joint action of electric field potential asymmetry and spin-orbit interaction. Over the years, the effect has been found in a variety of classes of materials experimentally and theoretically [61-64]. The mirror symmetry break in Janus transition metal dichalcogenide induces an intrinsic out-of-plane polarization, which stabilizes the spin nondegenerate states at the $\Gamma$ point, causing a powerful Rashba effect. [61]

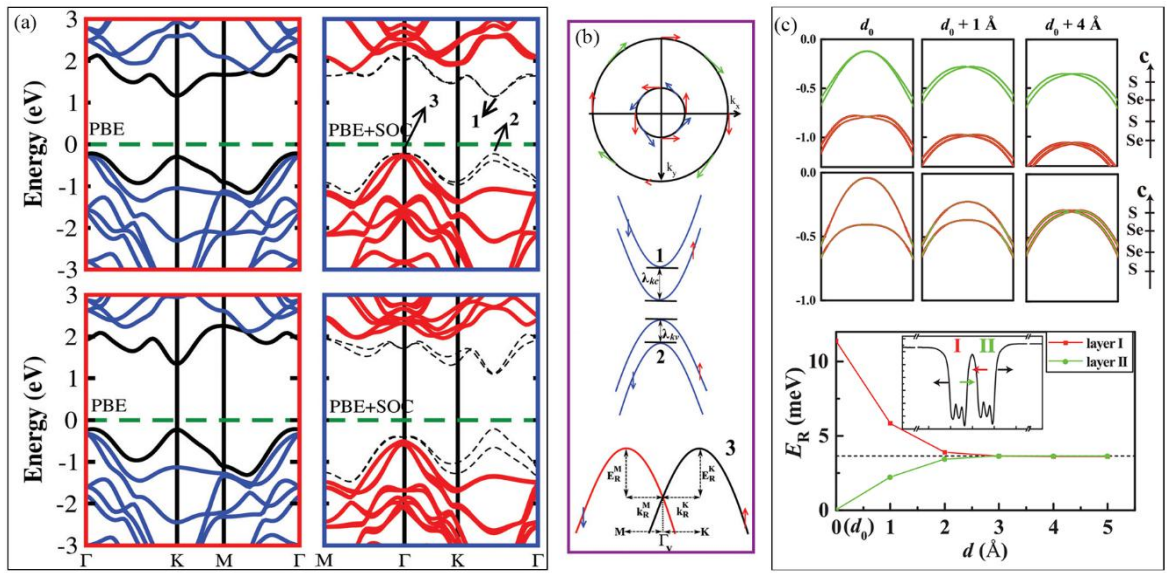

**Fig. 4** (a) Band structures of MoSSe–ZnO (up) and WSSe–ZnO (down) vdW heterostructures. (b) Graphics of the spin texture and the magnified views of the valence/conduction band-splitting at the K-point marked with the numbers of 1~3. (c) Top of valence band around $\Gamma$ of

WSSe bilayers with different interlayer distance and stacking orders (up), and the relationship between interlayer distance and the Rashba splitting energy of WSSe bilayer (down). The splitting energy of the WSSe monolayer is shown with the black dashed line. (a), (b) Reproduced from Ref.[42]. (c) Reproduced from Ref.[65].

In 2D Janus vdW heterostructures, depending on the substrate and the relative polarization direction of the Janus layer, the spin splitting and Rashba effect become obviously different. Take ZnO–WSSe and ZnO–MoSSe vdW heterostructures as the examples, Rashba spin splitting and band spin splitting for both configurations could be observed in the band structure [42]. The band spin texture around $\Gamma$, caused by the spin–orbit interactions, is displayed in **Fig. 4a** and **4b**. Due to the presence of the polarization, the Rashba splitting at valance and conduction bands are dependent on not only the stacking pattern, but also the choice of the substrate. The Rashba splitting at the valance band of MoSSe-ZnO is obviously weaker than that of WSSe-ZnO. Similarly, the same band spin split also exists in GeC-MoSSe and GeC-WSSe vdW heterostructures [44]. Particularly, the strength of the Rashba spin polarization could be tuned by adjusting the stacking patterns [44]. Besides, the change of interlayer distances also could tune the Rashba effects because of the competition between the interlayer and intralayer electric fields, see the examples of MoSSe and WSSe bilayer in **Fig. 4c** [65, 66]. The tunable Rashba spin polarization in 2D Janus vdW heterostructures provids a well platform for developing 2D spintronic devices [44].

### 3.2. Interface contact

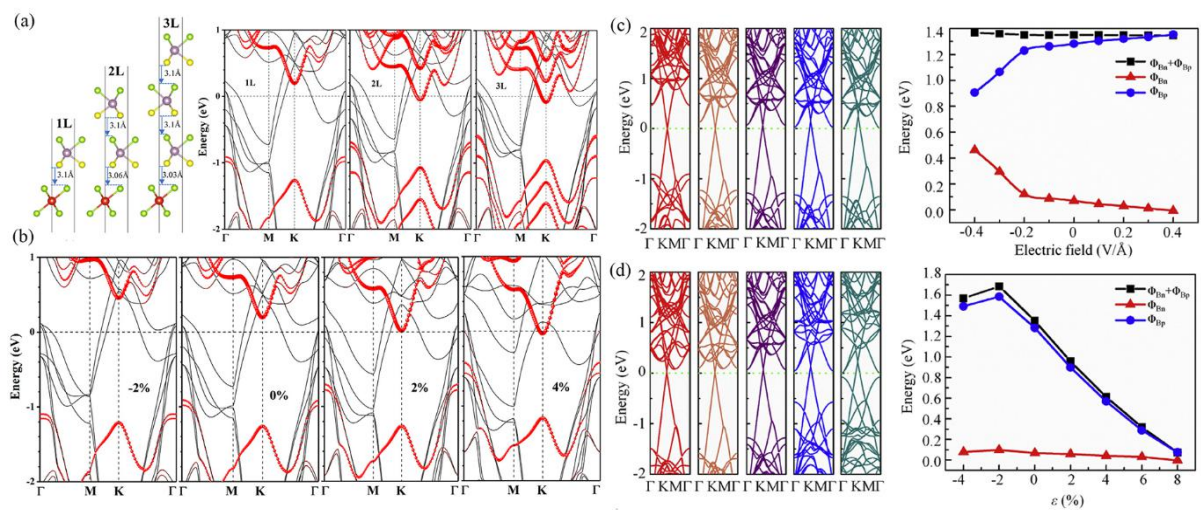

**Fig. 5** (a) Configurations and band structure of VSe$_2$-monolayer (1L)/bilayer (2L)/trilayer (3L) MoSSe vdW heterojunction. (b) Band structure of VSe$_2$-monolayer MoSSe vdW heterojunction under different biaxial strains. Band structures (left) and barriers (right) of the graphene-MoSSe vertical heterostructure under different (c) biaxial strains and (d) electric fields. (a), (b) Reproduced from Ref.[51]. (c), (d) Reproduced from Ref.[67].

The interface interaction plays a vital role on the performances of photoelectric nanodevices, such as photodetector and solar cell. A high barrier height will bring in a high resistance, preventing efficient electron injection and extraction. The device performance is therefore affected. Instead, to obtain the low resistance and high response rate in the heterostructure devices, the low barrier contact is desirable. Theoretically, the contact type can be distinguished by comparing the energy difference between the semiconductor band edges and the Fermi level at the interfaces. Specifically, the p-type barrier ($\Phi_{Bp}$) and n-type barrier ($\Phi_{Bn}$) are defined as following formulas:

$$\begin{cases} \Phi_{Bp} = E_F - E_{VBM} \\ \Phi_{Bn} = E_{CBM} - E_F \end{cases}$$

where $E_F$, $E_{VBM}$, and $E_{CBM}$ are the energy of Fermi level, VBM, and CBM, respectively. For the 2D Janus vdW heterojunction with appropriate vertical stacking pattern, low barrier contact has been found in the graphene-WSeTe [68], VSe$_2$-multilayer MoSSe [51], Ti$_2$C(OH)$_2$-MoSSe [69], Hf$_2$NF$_2$-MoSSe [70], Hf$_2$NF$_2$-WSSe [70], Hf$_2$N(OH)$_2$-MoSSe [70], Hf$_2$N(OH)$_2$-WSSe [70], germanene-bilayer MoSSe [71], graphene-trilayer PtSSe [72], and Mo$_2$C(OH)$_2$-MoSSe [69].

Furthermore, the barrier height and contact type can be controlled by appropriate vertical external strains (through adjusting the interlayer distance) and electric field, and the thickness of Janus 2D materials. In graphene-WSeTe heterostructure with the stacking order of G-Se-W-Te, under the negative electric field or tensile strain, $\Phi_{Bp}$ increases and $\Phi_{Bn}$ decreases. When the negative electric field is strong enough (>0.2 V/Å), the intrinsic p-type contact turns into the n-type contact. On the other hand, under the positive electric field or compressive strain, $\Phi_{Bp}$ decreases and $\Phi_{Bn}$ increases [68]. At the Graphen-MoSSe heterostructure interface, as demonstrated in **Fig. 5c** and **5d**, the vertical tensile strains (increasing the interlayer distance) reduce the p-type barrier height, so is the negative electric field [73]. Under compressive strain, the p-type barrier height of Hf$_2$NO$_2$-MoSSe and Hf$_2$NO$_2$-WSSe heterostructures drops [70]. Similar change in barrier height occurs in VSe$_2$-MoSSe under strain, as shown in **Fig. 5b** [51]. Besides, the thickness increase of Janus materials (MoSSe and PtSSe), could reduce the barrier height as well, see **Fig. 5a** [51, 72].

These theoretical findings, on the controllable barrier height and contact type, demonstrate the feasibility to design high-efficiency field-effect transistors with 2D Janus vdW heterojunctions.

Apart from contract barrier, tunnelling barrier is another important index to evaluate the performance of heterostructure based photoelectric nanosdevices. Generally, the tunnelling probability ($T_B$) is calculated by replacing the irregular real barriers with square barriers according to the WKB formula [69, 74]:

$$T_B = exp\left(-2\frac{\sqrt{2m\Delta_V}}{\hbar} \times w_B\right)$$

where $m$, $\hbar$, $\Delta_V$, and $w_B$ are respectively the mass of free electrons, reduced Planck's constant, and the height and width of the assumed square potential barrier. For the MoSSe-MXene ($Ti_2C(OH)_2$, $Mo_2C(OH)_2$, and $Sc_2N(O)_2$) vdW heterostructures, due to the weak interlayer interactions, the tunnelling probability is small (6.72%, 4.69%, and 11.65% for MoSSe-$Ti_2C(OH)_2$, MoSSe-$Mo_2C(OH)_2$, MoSSe-$Sc_2N(O)_2$, respectively) [69]. As for the case of $VSe_2$-multilayer MoSSe, the thickness increase of MoSSe layer has a crucial influence on raising the tunnelling barrier probability [51].

## 3.3. Optical properties

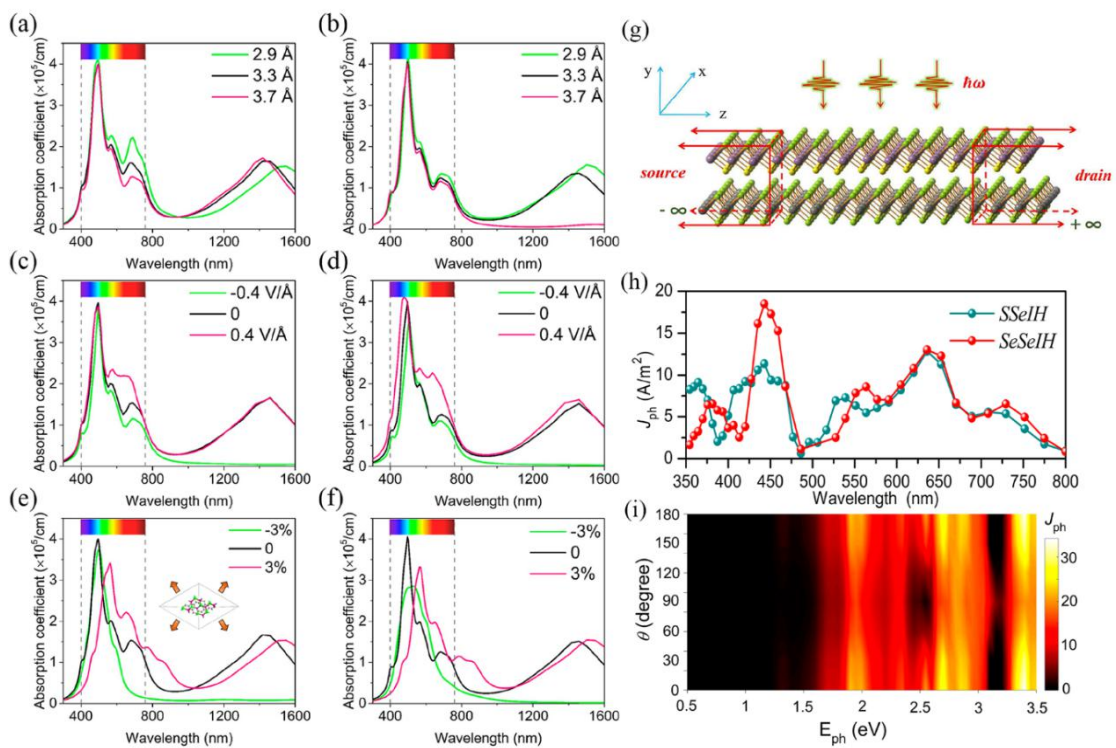

**Fig. 6** ((a), (b)) Interlayer distance, ((c), (d)) external electric field, and ((e), (f)) strain effects on the optical absorption coefficient for graphene-MoSSe vdW heterojunction with graphene-Se ((a), (c), and (e)) and graphene-S ((b), (d), and (f)) stacking order. (g) Schematic plot of the two-probe device used for photoinduced current calculations. Calculated photocurrents as a function of wavelength (h), and photon energy and polarizing angle (i) for vertical $MoS_2$-$WSe_2$ heterojunction. (a)~(f) Reproduced from Ref.[75]. (g)~(i) Reproduced from Ref.[41].

2D vdW heterojunction is a promising candidate for solar cell, photodetector, and photocatalyst [76-78]. For example, Zhang *et al.* found the combined advantages in the $MoS_2$-graphene heterostructure, namely broadband response and ultrafast relaxation of graphene and strong light-matter interaction of $MoS_2$ [76]. Compared with single $MoS_2$ and graphene, the heterostructure possesses superior photoresponse activities, which are mainly induced by the extremely efficient charge separation, strong light-matter interaction, and enhanced light absorption [77]. Similar phenomena and application were also revealed in the graphene-BlackP (black phosphorene) heterojunction [78].

2D Janus vdW heterostructures have also been predicted to possess wide absorption wavelength and high optical absorption coefficient in the visible regions. For example, in the MoSSe-GaN and MoSSe-AlN vdW heterostructures, the main peaks of the optical absorption reach $2.74 \times 10^5$ and $1.86 \times 10^5$ $cm^{-1}$ (at 425 and 536 nm), and $3.95 \times 10^5$ and $2.05 \times 10^5$ $cm^{-1}$ (at 412 and 528 nm), respectively [46]. The MoSSe-ZnO vdW heterostructure has a high absorption peak ($>10^5$ $cm^{-1}$) at 549.9 nm, and its optical absorption spectrum involves almost all the incident solar spectrum [79]. So are the MoSSe-SiC and WSSe-SiC vdW heterostructures [54]. In visible-light region, the main absorption peak of $GaS_{0.5}Se_{0.5}$-Arsenene heterostructure exceeds $10^5$ $cm^{-1}$, which is stronger than the one of isolated $GaS_{0.5}Se_{0.5}$ monolayer. The enhanced absorption coefficient is due to the reduced bandgap after the forming vdW heterostructure [49]. The Janus MoSSe-GaN heterobilayer has extensive light absorption spectrum (from visible light to ultraviolet light), while the major optical absorption peak is up to $10^5$ $cm^{-1}$ [53]. In the visible light, the chief absorption peak of graphene-WSeTe heterostructure reaches $5 \times 10^4$ $cm^{-1}$, twice as the value of Janus WSeTe monolayer. What's more, in ultraviolet region, its absorption coefficient even could be up to $10^5$ $cm^{-1}$ [68].

Moreover, the optical absorption coefficients could be effectively tuned by adjusting interlayer distance, external electric field and strains, see **Fig. 6a~6f**. In the MoSSe-ZnO vdW heterostructure, new adsorption peaks can be induced by strain in the visible light region [79]. For MoSSe-SiC and WSSe-SiC vdW heterostructures, it was found that the optical absorbance can be enhanced by strain in the visible light by emerging absorption peaks, leading to the blue-shift of optical absorption spectrum [54]. The effects of the interlayer distance and electric field was demonstrated in the graphene-MoSSe heterostructure. Specifically, the stronger interlayer coupling (realized by reducing the interlayer distance) results in a higher optical absorption coefficient. Meanwhile, the in-plane strains can render the absorption spectrum red-shift, and the optical absorption coefficient rises under the electric field of 0.4 V/Å. [75]

In addition, the photo-responsivity ($R_{ph}$), is another important parameter to scale the optical performances, which could be evaluated by following equation:

$$R_{ph} = \frac{J_{ph}}{eF_{ph}}$$

where $F_{ph}$ means the photon flux defined as the number of photons per unit time per unit area [80], and $J_{ph}$ is the photoinduced current. Employing a two-probe model under illumination (see **Fig. 6g**), the photoinduced current $J_{ph}$ can be obtained as:

$$J_{ph} = \frac{e}{\hbar} \int \frac{dE}{2\pi} \sum_{\alpha} T_{\alpha}(E)$$

where $\alpha$ and $T_{\alpha}(E)$ represent the lead electrode and the effective transmission coefficient, respectively. The effective transmission coefficient $T_{\alpha}(E)$ can be calculated as:

$$T_{\alpha}(E) = \text{Tr}\left\{i\Gamma_{\alpha}\left[(1-f_{\alpha})G_{ph}^{<} + f_{\alpha}G_{ph}^{>}\right]\right\}$$

where $\Gamma_{\alpha}$, $f_{\alpha}$ and $G_{ph}^{>/<}$ denote the line-width function, Fermi function, and greater/lesser Green's function including electron-photon interactions [81-83]. With the external bias voltage (0.2 V) and standard incident light power density (1 kW·m$^{-2}$), it has been found that the MoSSe-WSe$_2$ heterostructure with suitable stacking pattern exhibits high photocurrents in a broad range of spectrum. As displayed in **Fig. 6h**, its maximum value reaches 0.017 A/W at 442 nm incident laser wavelength, which is comparable with those of lateral InSe-InTe heterostructure (0.030 A/W) [84] and vertical MoS$_2$-WSe$_2$ heterojunction (0.011 A/W) [85]. In addition, according to the calculated photocurrents at different polarization angle θ, shown in **Fig. 6i**, this heterostructure was found to show higher photo-responsivity at θ = 0° and 180°.

Additionally, the distinct properties of exciton render 2D Janus vdW heterostructure promising for optoelectronic and valleytronic devices. Based on the electronic and optical structures, it is found that, there is a drastic competition between intralayer and interlayer excitons in the MoSSe-WSe$_2$ heterostructure. The built-in electric field, caused by intrinsic dipole of Janus MoSSe monolayer, can affect exciton significantly. In the case of S-Mo-Se-Se-W-Se stacking configuration, due to the weak interlayer coupling, the intralayer exciton plays a dominant role in exciton. Whereas, in Se-Mo-S-Se-W-Se stacking order, the powerful interlayer coupling induces a coherence cancellation at the edge of valence band, and finally causes a transition of bright-to-dark exciton [86]. Moreover, with better electron screening, the MoSSe-C$_3$N$_4$ and MoSTe-C$_3$N$_4$ heterostructures separately have a lower exciton binding energy than the isolated MoSSe and MoSTe monolayer, indicating an easier carrier separation [87].

## 4. Potential applications

### 4.1. Energy harvest

**Photocatalytic overall water-splitting**

Because of the rising energy demands, it is essential to search for renewable energy sources. Converting clean solar energy into chemical energy is considered as the most promising technology, which has received intensive interdisciplinary attentions [59, 88, 89]. Until now, a number of photocatalytic materials has been developed, such as TiO$_2$ [90], ZnO [91], SrTiO$_3$ [92], g-C$_3$N$_4$ [3], and so on. Unfortunately, a single phase semiconductor often suffers from several shortcomings including low quantum efficiency, high charge recombination, and serious chemical back-reactions [12, 13]. It is generally believed that the heterostructure, which combines different materials and associated properties together, is a feasible solution to overcome these problems [93, 94]. Hence, heterostructure devices are developed to increase the solar energy conversion efficiency and chemical activities [95, 96]. Due to the well separation of photogenerated electron-hole pairs, good ability of optical absorption, appropriate band alignments, and excellent carrier mobility, many 2D Janus vdW heterojunctions have been designed for the photocatalytic overall water-splitting application, such as ZnO-MoSSe [42], ZnO-WSSe [42], GeC-WSSe [44], GeC-MoSSe [44], MoSSe-GaN

[46, 56], MoSSe-AlN [46, 56], GaS$_{0.5}$Se$_{0.5}$-Arsenene [49], MoSSe-SiC [54], WSSe-SiC [54], MoSSe-WSSe [45], MoSSe-C$_3$N$_4$ [87], and MoSTe-C$_3$N$_4$ heterostructures [87].

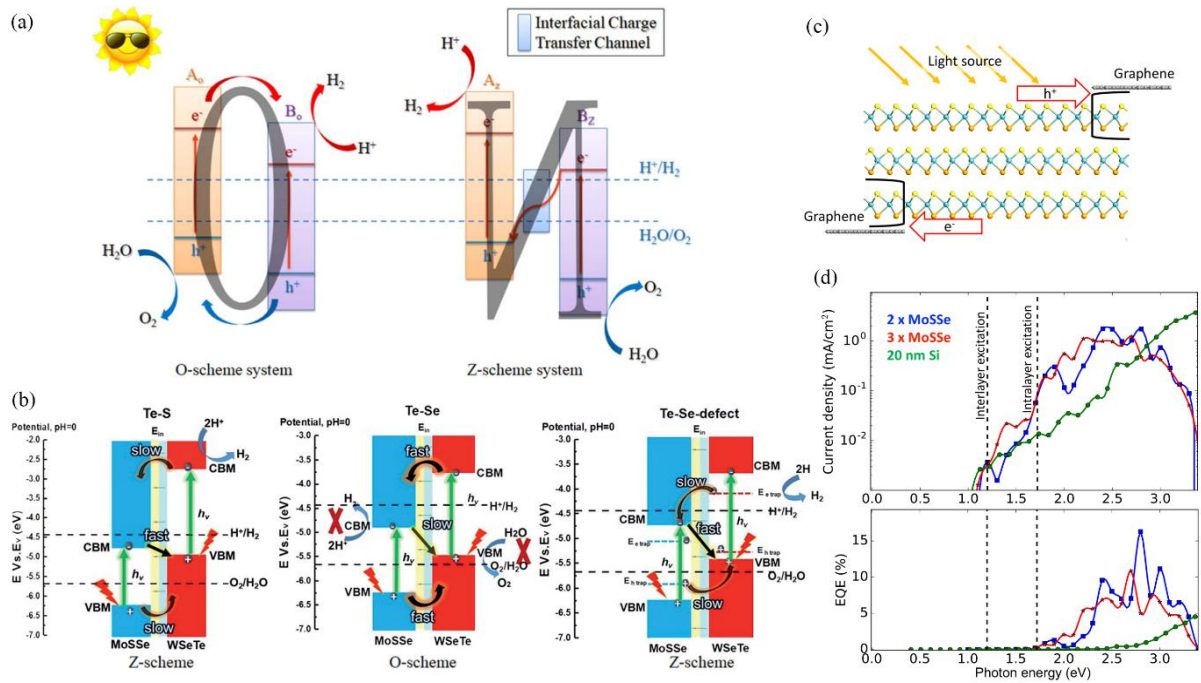

**Fig. 7** (a) Schematic illustrations of the O-scheme (left) and Z-scheme (right) systems. (b) Schematic diagram of the charge transfer mechanism and band edge position of MoSSe-WSeTe vertical heterostructure with Te–S, Te–Se and Te–Se-defect stacking patterns. (c) Trilayer Janus MoSSe device structure used to calculate transmission. (d) Photocurrent density (up) and external quantum efficiency (EQE) (down) for the trilayer MoSSe structure, shown in **Fig. 7c**, compared to the two layer device and a 20 nm silicon thin film device. (a) Reproduced from Ref.[74]. (b) Reproduced from Ref.[97]. (c), (d) Reproduced from Ref.[98].

As is well known, the type-Ⅱ band alignment in vertical heterojunction is beneficial to improve the charge spatial separation [99, 100]. As shown in **Fig. 7a**, according to the transfer pathway of photogenerated charge, these vdW heterostructures could be majorly divided into two categories: O-scheme and Z-scheme systems [74]. The O-scheme system could save photogenerated carriers, having as many carriers as possible participate in the redox reactions. However, the redox ability of the carriers in this system will be weakened, which is unfavourable for efficient photocatalytic overall water-splitting. In contrast, in Z-scheme systems, the highest redox ability of the photogenerated carriers among the components can be retained and utilized. Hence, the components for the reduction and oxidation do not have to be suitable for overall water splitting themselves. Up to now, lots of efforts have been devoted to this filed [18, 101-105]. The Z-scheme system based on 2D Janus vdW heterojunction has also been designed for photocatalytic hydrogen production. For example, on the basics of nonadiabatic molecular dynamics calculations, the MoSSe-WSeTe has been verified to be a potential direct Z-scheme photocatalyst for hydrogen evolution reaction. Through tuning the time difference between photo-excited carrier transfer and recombination at the interface, the stacking configurations, as well as the surface chalcogen vacancies could switch the charge transfer path from Z-scheme to O-scheme, as exhibited in **Fig. 7b**.[97]

Besides the band alignment, the interlayer built-in electric field induced by charge redistribution at the interface also affects the photocatalytic efficiency by tuning the carriers' spatial separation [18, 74]. Different from the normal vdW heterojunction, in the 2D Janus vdW heterostructures, beside the interlayer built-in electric field, there is an intrinsic intralayer polarization in the Janus layer, which could further promote the spatial separation of carriers at the interlayer, enabling asymmetric carrier tunnelling through all layers. As a result, in contrast of the photocurrent decreasing with number of layers in classical stacked transition metal dichalcogenides [106], as displayed in **Fig. 7d**, the generated photocurrent nearly is independent of the thickness in graphene-multilayer MoSSe vdW heterostructures [98].

**Piezoelectric, thermoelectric, and photovoltaic applications**

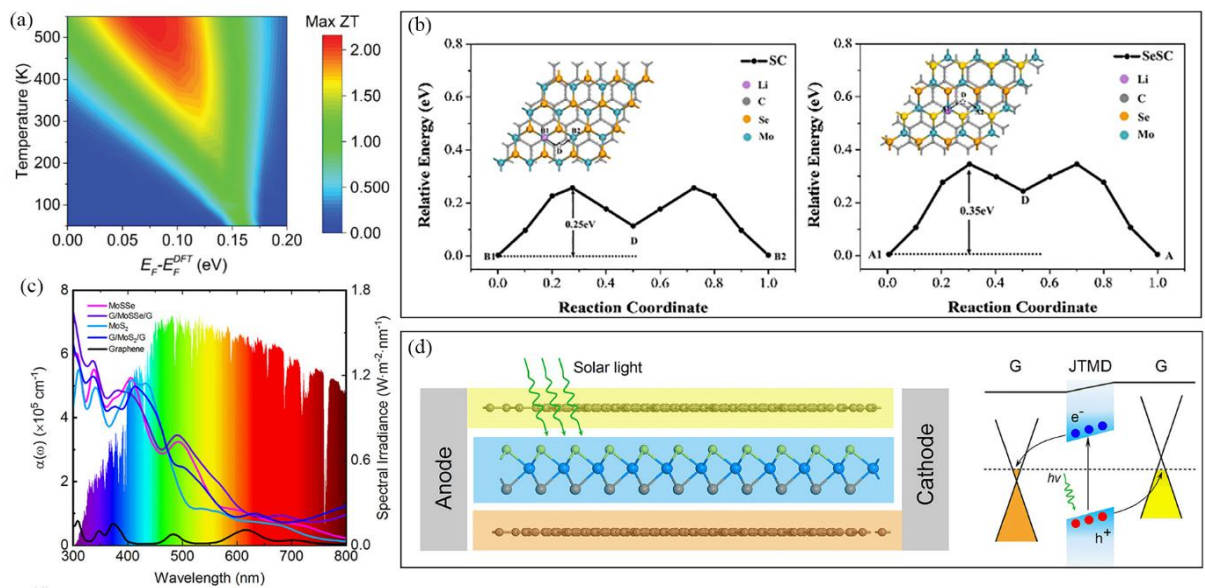

**Fig. 8** (a) The max ZT variation of the graphene-MoSSe heterostructure nanoribbon with different temperature and chemical potential ($E_F - E_F^{DFT}$). (b) The best diffusion pathway and the energy barrier of the Li atom diffusing on graphene-MoSSe heterostructure with Se-Mo-S-C (left) and S-Mo-Se-C (right) stacking orders. (c) Light absorption coefficient of the MoSSe monolayer, $MoS_2$ monolayer, graphene, graphene-MoSSe-graphene vdW heterostructure, and graphene-$MoS_2$-graphene vdW heterostructure. (d) Photovoltaic device designed with graphene-MoSSe/MoSeTe-graphene vdW heterostructure (left) and schematic band diagrams for its carrier separation (right). The band in Janus transition metal dichalcogenides layer slopes downward from the heavy chalcogen atoms (Se and Te) side to the lighter chalcogen atom (S and Se) side. (a) Reproduced from Ref.[107]. (b) Reproduced from Ref.[108]. (c), (d) Reproduced from Ref.[109].

Besides converting solar energy into chemical energy through photocatalytic reaction, converting mechanical and thermal energy into electric energy are also important paths to harvest energy. Although the relevant results are still rare, some of 2D Janus vdW heterojunctions have been demonstrated to have potential applications for piezoelectric, thermoelectric, and photovoltaic devices. For example, the MoSSe-BlueP vdW heterostructure is found to be dynamically stable with enhanced elastic moduli. Its out-of-plane piezoelectric response, characterized by the piezoelectric coefficient $e_{311}$ (0.081 C/m), is stronger than that of Janus MoSSe monolayer (0.058 C/m), indicating the potential in piezoelectric application [47]. In graphene-MoSSe stacked nanoribbons, the thermal merit (ZT) values can reach up to 2.01 at 300 K, which is significantly larger than the intrinsic value of graphene (ZT=0.05). As displayed in **Fig. 8a**, the maximum ZT values increase among the low temperature range (≤350K) but decrease among the high temperature range (>350K). The significantly enhanced ZT values indicates the potential application in thermoelectric devices [107]. The carrier dynamics show that, in graphene and 2D MoSSe/MoSeTe vdW sandwich heterojunction (graphene-MoSSe-graphene, graphene-MoSeTe-graphene), photogenerated carriers transfer from MoSSe/MoSeTe to different graphene layers with a preferred direction (shown in **Fig. 8d**) on the time scale of hundreds of femtoseconds. The asymmetry potential, induced by the intrinsic built-in electric field in MoSSe/MoSeTe, promotes the migration. More specifically, the photogenerated holes (electrons) could flow to the graphene layers at the Se side with a higher (lower) potential, at the same time, the same transfer of the photogenerated electrons (holes) is suppressed. The high degree of separation for the photogenerated carrier, combined with the strong visible light absorption (larger than $10^5$ com$^{-1}$, see **Fig. 8c**), make these vdW sandwich heterojunction suitable for photovoltaic cells with a high power conversion efficiency [109].

## 4.2. Optoelectronic devices

To be an excellent photoelectric device, the high photo-responsivity is essential, which mainly depend on the photoexcitation dynamics and excitonic effects. Janus vdW heterojunctions, with type-Ⅱ band alignment, have been designed to increase exciton and charge dissociation, such as the MoSSe-WSe$_2$ vertical heterostructure. Both the photo-response and absorption coefficients of the heterojunction are found to have optical activity in a broad range of the visible spectrum [41, 110]. Because of the suitable band gap, fast charge separation, and slow electron-hole recombination, the MoSSe-WSe$_2$ heterostructure shows a remarkable optoelectronic performance (up to 0.017 A/W at 442 nm) [41]. In MoSSe-graphene vdW heterostructure, the doping level and optical plasmon energies of graphene can be significantly influenced by the intrinsic vertical electric dipole and the thickness of Janus layers. Furthermore, their optical plasmon energies [111] and optical absorption coefficient [75] can also be adjusted by choosing different Janus materials. In addition, compared with the free Janus WSeTe and graphene monolayer, the graphene-WSeTe heterostructure has enhanced optical absorption. In visible region, the absorption coefficient reaches $5 \times 10^4$ cm$^{-1}$, twice the one of Janus WSeTe monolayer. Meanwhile, in the ultraviolet region, the absorption coefficient achieves up to $10^5$ cm$^{-1}$ [68]. These theoretical findings indicate that the vertical Janus vdW heterojunctions have promising potential in optoelectronic devices.

## 4.3. Li-ion batteries

The ion battery, as the fastest growing electrochemical cell, is considered to be a significant improvement over the traditional nickel-cadmium batteries, and has been widely applied in wireless portable electronic products. Especially, the lithium-ion battery revolution brings great convenience to our life. Layered MoS$_2$ with a high theoretical specific capacity (669 mAh/g), has attracted extensive attention for ion batteries [112]. Unfortunately, the low intrinsic electrical conductivity results in some adverse factors, such as fast capacity fading, low rate behaviour, and sluggish dynamics, which, to a great extent, limits the application of layered MoS$_2$ in ion batteries [113]. Excitingly, Janus MoSSe layer with an intrinsic dipole is theoretically predicted as a potential electrode material. The intrinsic dipole in Janus MoSSe layers can make the Li-ions stably adsorb on the side of the S layer and enable more Li-ions stored. With the suitable open circuit voltage (0.62 V for the monolayer and 1.01 V for the bilayer) *vs.* Li$^+$/Li, the theoretical capacities even could achieve 776.5 and 452.9 mAh/g for the monolayer and bilayer, respectively. Besides, the good electrical conduction and the small Li-ion/Li-vacancy migration barrier (≤0.34 eV) declare a fast Li-ion diffusion [113]. In addition, it has been found that, the coverage of graphene makes the adsorption of lithium atoms more stable. And, as shown in **Fig. 8b**, the diffusion barriers on the surface (MoSSe side) of the MoSSe-graphene vdW heterostructure, (≤0.34 eV) are comparable to the one on the bare MoSSe monolayer. Its maximum lithium storage capacity reaches up to 390 mAh/g, and the open circuit voltage range is suitable for the utilization as an anode material [108].

## 5. Conclusion and outlook

In summary, here we reviewed the recent research achievements from the fundamental properties to potential applications of 2D Janus vdW heterojunctions. It is found that, due to the unique structures and complex intralayer-interlayer polarization interaction, these heterojunctions could exhibit novel physicochemical properties, such as tuneable band gap and band edge positions, wide light adsorption spectra, controllable contact resistance, and sufficient redox potential. These excellent electronic and optical properties render the 2D Janus vdW heterostructures promising to be used in energy conversion and electronics devices.

Even so, the study for 2D Janus vdW heterojunction is still at an initial stage, with many challenges and opportunities. First of all, most of these investigations are from theoretical simulations, yet to be confirmed by experiments. To verify these outstanding performances forecasted by theoretical simulations, the highest priority is to fabricate this vdW heterojunction in reality. For achieving this aim, high-quality 2D Janus materials need to be synthesised first. However, up to now, due to the harsh synthesis conditions (strict requirements for the temperature and pressure), only Janus MoSSe and WSSe monolayers have been synthesised successfully. Especially, the Janus MoSSe monolayer prepared with the CVD method has significant levels of defects, which cause the weak photoluminescence and crack formation [26]. Hence, generalizing the synthetic methods to fabricate new Janus layer materials and developing user-friendly fabricating strategies to improve the sample quality are eagerly demanded. Furthermore, how to exfoliate the Janus layer materials from their substrates and stack them on other 2D materials to form the heterostructures, and how to directly grow the Janus layer materials on the desired substrate are both challenges, which need to be solved urgently.

On the other hand, deeper understandings for several aspects of 2D Janus vdW heterostructures are essential to their practical application. For example, some important photocatalytic properties, such as the exciton binding energy, solar-to-hydrogen (STH) efficiency and photoexcited carriers driving force [114, 115], have been intensively investigated for normal vdW heterojunctions with the state-of-the-art theoretical computing technique. However, for the 2D Janus vdW heterostructure, the comprehensive investigations for these properties are still absent. In addition, the interface contact and stacking have been studied as reviewed above, but the accurate description on the interlayer charge transfer path is an unsolved problem. Generally, this description is based on the direction of the interlayer built-in electric field and the difference between charge transfer and recombination time [17, 18, 97]. For 2D Janus vdW heterojunction, the influence of intralayer electric fields direction switch (antiparallel and parallel to the direction of interlayer electric fields) on the charge space-distribution may be an additional factor on judging the path of interlayer charge transfer, which needs numerous works to verify in the future. Furthermore, the stability of semiconductors under illumination in the aqueous solution is an important factor in the practical photocatalytic applications. Although some experiments have proved that vdW heterostructure can work steadily for photocatalytic water-splitting reaction, up to now, neither experimental nor theoretical study on this operation stability of 2D Janus vdW heterojunctions could be found. Last but not the least, all the investigations in spintronics, field-effect transistors, Li-ion batteries and optoelectronic application are quite fundamental, which provides numerous opportunities but also challenges for the researchers in related fields.

**Acknowledgment**

This work is supported by National Natural Science foundation of China (Grants No. 11804006), Henan Key Program of Technology Research and Development (No. 182102310907), and Henan College Key Research Project (No. 19A430006). L.J. gratefully acknowledges China scholarship council for its support (No. 201908410036). L.K. gratefully acknowledges financial support by the ARC Discovery Project (DP190101607).